\documentclass[twocolumn,twoside,slac_two]{revtex4}
\usepackage{graphicx}
\usepackage{fancyhdr}
\begin{document}

\title{X-ray Emission from Middle-Aged Gamma-Ray Pulsars}

%

\author{S. Kisaka}
\affiliation{Institute of Particle and Nuclear Studies, KEK, Tsukuba, Ibaraki, 305-0801 Japan}
\author{S. J. Tanaka}
\affiliation{Institute for Cosmic Ray Research, University of Tokyo, Kashiwa, Chiba, 277-8582 Japan}

\begin{abstract}
Electrons/positrons produced in a pulsar magnetosphere emit synchrotron radiation, which is widely believed as the origin of the non-thermal X-ray emission detected from pulsars. 
Particles are produced by curvature photons emitted from accelerated particles in the magnetosphere.
These curvature photons are detected as pulsed $\gamma$-ray emissions from pulsars with age $\lesssim10^6$ yr. 
Using $\gamma$-ray observations and analytical model, we impose severe constraints on the synchrotron radiation as a mechanism of the non-thermal X-ray emission. 
In most middle-aged pulsars ($\sim10^5-10^6$ yr) which photon-photon pair production is less efficient in their magnetosphere, we find that the synchrotron radiation model is difficult to explain the observed non-thermal X-ray emission.

\end{abstract}

\maketitle

\thispagestyle{fancy}


\section{INTRODUCTION}

Pulsed non-thermal X-ray emissions are detected from rotation-powered pulsars.  
Synchrotron radiation is widely believed as the emission mechanism \cite[][]{TCS08}.
Electrons and/or positrons produced in the magnetosphere initially have non-zero value of the pitch angle, so that they emit the synchrotron radiation. 
Thus, the non-thermal X-ray emission is important to clarify the particle production in the pulsar magnetosphere.

Observed non-thermal X-ray luminosity $L_{\rm nth}$ is typically $\sim10^{-3}-10^{-5}$ times smaller than the spin-down luminosity \cite[e.g., ][]{Kar+12}.
Non-thermal emission is detected at $\nu_{\rm obs}\gtrsim1$ keV. At soft X-ray band ($\lesssim1$ keV), thermal component significantly contributes to the total luminosity. 
The origin of this thermal luminosity is considered as the bombardment of particles moving to the polar cap surface \cite[e.g., ][]{HR93}.
The luminosity ratio between the non-thermal and the thermal components is typically $\xi\equiv L_{\rm nth}/L_{\rm th}\sim10^{-1}-10$ \cite[e.g., ][]{MDC11}. 

In the magnetosphere of older pulsars, the pair production through the photon-photon collision is less effective.
As a pulsar gets old, the spin period $P$ increases as well as the radius of the light cylinder $R_{\rm lc}=Pc/2\pi$ increases, where $c$ is the speed of light.
For the pulsars with age $\gtrsim10^5$ yr, the luminosity of the whole surface thermal emission significantly decreases \cite[e.g., ][]{YP04}.
Then, the number density of the X-ray photons at the outer magnetosphere of pulsars with $\gtrsim10^5$ yr is too small to produce the significant number of pairs through photon-photon collision \cite[e.g., ][]{TC09, H13}.

Magnetic pair production is considered as one of the main pair-production process in the magnetosphere \cite[e.g., ][]{S71}.
Some authors \cite[e.g., ][]{CZ99, ZC97, CGZ98} propose that the synchrotron radiation from pairs produced through the magnetic pair-production explains the non-thermal X-ray emission from pulsars including middle-aged one. 
These pairs are produced from curvature photons emitted by accelerated particles with inward direction.

Recently, Kisaka \& Tanaka \cite{KT14} argue that the synchrotron radiation model with ingoing accelerated particles and magnetic pair production does not explain the observed non-thermal emission for pulsars with $\gtrsim10^6$ yr (Figure 4 in \cite{KT14}).
Middle-aged pulsars locate the allowed region on $P-\dot{P}$ diagram in their results.

In the analysis of Kisaka \& Tanaka \cite{KT14}, one of the most important parameter is the Lorentz factor of the accelerated particles. 
Since there is no observational constraint on the Lorentz factor in old pulsars, Kisaka \& Tanaka \cite{KT14} adopt the maximum value (equation 2 in \cite{KT14}). 
This value is much larger than the realistic one, which is determined by the force balance between the electric field acceleration and the radiation reaction force \cite[e.g., ][]{CHR86}. 
In the model of \cite{KT14}, smaller Lorentz factor of acceleration particles always more stringent limits on synchrotron radiation model for the non-thermal X-ray emission.

{\it Fermi} detects the pulsed $\gamma$-ray emission from more than 100 pulsars including middle-aged ones \cite{catalog}. 
The cutoff energies of detected pulsars are typically $\sim1$ GeV. 
Observed $\gamma$-ray emission is considered as the curvature radiation from accelerated particles. 
Then, we can evaluate the Lorentz factor of the accelerated particles from the characteristic energy of the curvature radiation.  
Therefore, $\gamma$-ray observations could impose more realistic and stringent constraints on the synchrotron radiation model.

In this proceeding, we give the constraints on the synchrotron radiation as the mechanism of the non-thermal X-ray emission from middle-aged gamma-ray pulsars. 
In particular, we impose the upper limit on the Lorentz factor of accelerated particles from $\gamma$-ray observations. 
In Sec. 2, we introduce some assumptions and two constraints for the location of the X-ray emission region. Results and discussion are presented in Sec. 3. 

\section{Constraints}

We investigate the case that produced particles in the magnetosphere emit synchrotron radiation in X-ray band. 
We only focus on the magnetic pair-production as the production mechanism of synchrotron emitting particles. 
The magnetic pair-production is efficient within $\lesssim3-5R_{\rm NS}$ for GeV $\gamma$-ray photons, where $R_{\rm NS}$ is the radius of the neutron star. 
Since no attenuation feature due to the magnetic pair-production is detected in the observed $\gamma$-ray spectra \cite{catalog} in $\gamma$-ray pulsars, the particle acceleration occurs at the relatively outer magnetosphere ($\gtrsim10R_{\rm NS}$) considered by such as outer gap model \cite[e.g., ][]{CHR86}. 
Therefore, we only consider that the accelerated particles move to the direction of the neutron star. 
For the structure of the magnetic field, we assume dipole field.

In our definitions, "primary particles" means the electrons and positrons that are accelerated and emit curvature photons that can convert pairs. 
"Secondary particles" means those produced outside the acceleration region of primary particles $r_{\rm pri}$, including the second and higher generation particles. 
The production and emission locations of second and higher generation particles $r_{\rm sec}$ are almost the same, and then we 
do not separately treat second and higher generation particles.

The difference from previous model \citep{KT14} is that an observed characteristic energy of the $\gamma$-ray emission $E_{\rm cur}$ imposes a constraint on the Lorentz factor of the primary particles $\gamma_{\rm p}$. 
The observed $\gamma$-ray emission at $\sim1$ GeV is considered as the curvature radiation from the primary particles \cite[e.g., ][]{DH96, CHR86}. 
The characteristic energy of curvature radiation is described by
\begin{eqnarray}\label{sec2:E_cur}
E_{\rm cur}=0.29\frac{3}{4\pi}\frac{h\gamma_{\rm p}^3c}{R_{\rm cur}(r_{\rm pri})},
\end{eqnarray}
where $h$ is Planck constant. From the assumption of the dipole magnetic field, we use the approximation $R_{\rm cur}(r_{\rm pri})\sim(r_{\rm pri}R_{\rm lc})^{1/2}$ as a curvature radius of a field line. 
Hereafter, we use $Q_x\equiv Q/10^x$ in cgs units, except for a frequency $h\nu_{\rm keV}\equiv h\nu/1$keV and an energy $E_{\rm cut,GeV}\equiv E_{\rm cur}/1$ GeV. 

Note that observed $\gamma$-ray photons are emitted from outgoing particles. 
The Lorentz factor of the outgoing particles tends to be larger than that of ingoing one as following reason. Because the magnetic field and ambient photon density is larger for smaller radial distance from a neutron star, the location of the particle production is near the inner boundary of the particle acceleration region \cite[e.g., ][]{TSHC06, T10, H13}. 
Then, the outgoing particles obtain larger energy due to the electric field acceleration \cite{TCS08}. 
Therefore, we consider that the Lorentz factor estimated from equation (\ref{sec2:E_cur}) is the upper limit on the ingoing accelerated particles.

The observed frequency of non-thermal component $\nu_{\rm obs,keV}\gtrsim1$ keV and the luminosity ratio $\xi\sim0.1-10$ impose the lower and upper limits on the emission location of synchrotron radiation \cite{KT14}. 
Following Kisaka \& Tanaka \cite{KT14} we consider two conditions, the characteristic frequency (Sec. \ref{cf}) and the luminosity of synchrotron radiation (Sec. \ref{nl}).

\subsection{Characteristic frequency}
\label{cf}

To emit the synchrotron radiation, particle momentum perpendicular to the magnetic field has to satisfy the condition $\gamma\sin\alpha\sim\gamma\alpha>1$, where $\gamma$ is the particle Lorentz factor and $\alpha\le1$ is the pitch angle which is typically much smaller than 1.
This condition gives a lower limit on the frequency of the synchrotron radiation,
\begin{eqnarray}\label{sec2:frequency}
\nu_{\rm obs} \gtrsim \frac{eB(r_{\rm sec})}{2\pi m_{\rm e}c\alpha},
\end{eqnarray}
where $e$ and $m_{\rm e}$ are the charge and the mass of a electron. 
Using the assumption of a dipole magnetic field, the strength of the magnetic field is $B(r_{\rm sec})\sim B_{\rm s}(r_{\rm sec}/R_{\rm NS})^{-3}$, where $B_{\rm s}$ is the magnetic field at the surface. 
Then, the condition (\ref{sec2:frequency}) gives the lower limit for the emission location \cite[e.g., ][]{OS70, RD99}, 
\begin{eqnarray}\label{sec4:r_ct}
r_{\rm ct,6}\sim2.9\alpha^{-1/3}\nu_{\rm obs,keV}^{-1/3}B_{\rm s,12}^{1/3}.
\end{eqnarray}

\subsection{Non-thermal luminosity}
\label{nl}

Observed luminosity of the non-thermal component $L_{\rm nth}$ imposes the limit on the emission location. 
The luminosity of the synchrotron radiation is described as $P_{\rm syn}N_{\rm s}$, where $P_{\rm syn}$ is the power of the synchrotron radiation emitted by a single electron and $N_{\rm s}$ is the number of the secondary particles.
In our model, secondary particles are produced by the curvature photons of the primary particles.
Then, the number of the secondary particles are described by $N_{\rm s}\sim N_{\gamma}\tau N_{\rm p}$, where $N_{\gamma}$ is the effective number of curvature photons emitted by a single primary electron, $\tau$ is the optical depth for the pair production and $N_{\rm p}$ is the number of the primary particles.
Considering the higher generation pairs, the number of produced particles is maximally increased by a factor of $\gamma_{\rm s,pair}(r_{\rm pri})/\gamma_{\rm s,lt}(r_{\rm sec})$, where $\gamma_{\rm s,pair}$ is the Lorentz factor of the particle produced by a curvature photon and $\gamma_{\rm lt}$ is the lower threshold value of the Lorentz factor for the magnetic pair production. 
Therefore, the required condition to explain the observed luminosity is described by
\begin{eqnarray}\label{sec2:number2}
P_{\rm syn}(r_{\rm sec})N_{\gamma}(r_{\rm pri},r_{\rm sec})\tau N_{\rm p}(r_{\rm sec}) \nonumber \\
\times\frac{\gamma_{\rm s,pair}(r_{\rm pri})}{\gamma_{\rm s,lt}(r_{\rm sec})}>L_{\rm nth}.
\end{eqnarray}

We evaluate the number of the secondary particles $N_{\rm s}\sim N_{\gamma}\tau N_{\rm p}$. 
The effective number of the curvature photons $N_{\gamma}(r_{\rm pri},r_{\rm sec})$ is
\begin{eqnarray}\label{sec2:N_gamma}
N_{\gamma}(r_{\rm pri}, r_{\rm sec})&\sim&\dot{N}_{\gamma}(r_{\rm pri})t_{\rm ad}(r_{\rm sec}) \nonumber \\
&\sim&\frac{P_{\rm cur}(r_{\rm pri})}{E_{\rm cur}}t_{\rm ad}(r_{\rm sec}),
\end{eqnarray}
where $P_{\rm cur}(r_{\rm pri})$ is the power of curvature radiation by a single electron,
\begin{eqnarray}\label{sec2:P_cur}
P_{\rm cur}(r_{\rm pri})=\frac{2e^2c}{3R_{\rm cur}^2(r_{\rm pri})}\gamma_{\rm p}^4.
\end{eqnarray}
The Lorentz factor of the primary particle $\gamma_{\rm p}$ is obtained by the observed energy $E_{\rm cur}$ (equation \ref{sec2:E_cur}).
In the derivation of equation (\ref{sec2:N_gamma}), we assume that the primary particles continuously emit the curvature radiation during the advection timescale of the secondary particles,
\begin{eqnarray}\label{sec2:t_ad}
t_{\rm ad}(r_{\rm sec})\sim \frac{r_{\rm sec}}{c}.
\end{eqnarray}
For the optical depth of magnetic pair production, we use 
\begin{eqnarray}\label{sec2:tau}
\tau\sim1
\end{eqnarray}
as long as the curvature photon energy exceeds the pair-production threshold for the magnetic pair-production \cite{RS75}
\begin{eqnarray}\label{sec2:threshold}
 \frac{E_{\rm cur}}{2m_{\rm e}c^2}\frac{B_{\perp}(r_{\rm sec})}{B_{\rm q}}>\frac{1}{15},
\end{eqnarray}
where $B_{\rm q}=m_{\rm e}^2c^3/e\hbar\sim4.4\times10^{13}{\rm G}$ and $B_{\perp}(r_{\rm sec})\sim B(r_{\rm sec})\alpha$. 
In our model, we consider the ingoing primary particles as the origin of pair cascade process. 
The kinetic energy flux of them $\dot{N}_{\rm p}\gamma_{\rm p}m_{\rm e}c^2$ is constrained by the observed thermal luminosity $L_{\rm th}$, 
\begin{eqnarray}\label{sec2:dotN_p,in}
\dot{N}_{\rm p}=\frac{L_{\rm th}}{\gamma_{\rm p}m_{\rm e}c^2}.
\end{eqnarray}
The number of the primary particles is described by 
\begin{eqnarray}\label{sec2:N_p}
N_{\rm p}\sim\dot{N}_{\rm p}t_{\rm cool},
\end{eqnarray}
because the cooling timescale of the secondary particle $t_{\rm cool}$ is always shorter than the advection timescale at the region where magnetic pair production occurs.
This cooling timescale is described by 
\begin{eqnarray}\label{sec2:t_cool}
t_{\rm cool}(r_{\rm sec})\sim
{\displaystyle \frac{\gamma_{\rm s,syn}(r_{\rm sec})\alpha m_{\rm e}c^2}{P_{\rm syn}(r_{\rm sec})}},
\end{eqnarray}
where the Lorentz factor of the secondary particles $\gamma_{\rm s,syn}$ is determined by the observed frequency,
\begin{eqnarray}
\nu_{\rm obs}=0.29\frac{3}{4\pi}\gamma_{\rm s,syn}^2(r_{\rm sec})\frac{eB(r_{\rm sec})\alpha}{m_{\rm e}c}.
\end{eqnarray}
The Lorentz factor $\gamma_{\rm s,pair}(r_{\rm pri})$ of secondary particles is
\begin{eqnarray}\label{sec2:gamma_s}
\gamma_{\rm s,pair}(r_{\rm pri})=\frac{E_{\rm cur}}{2m_{\rm e}c^2}.
\end{eqnarray}
From equation (\ref{sec2:threshold}), we take the threshold th eLorentz factor as
\begin{eqnarray}\label{sec2:gamma_s,lt}
\gamma_{\rm s,lt}(r_{\rm sec})=\frac{1}{15}\frac{B_{\rm q}}{B_{\perp}(r_{\rm sec})}.
\end{eqnarray} 

The relation between two point $r_{\rm pri}$ and $r_{\rm sec}$ is geometrically given by (Appendix in \cite{KT14})
\begin{eqnarray}\label{sec4:r_pri}
r_{\rm pri,6}\sim27r_{\rm sec,6}^{2/3}R_{\rm open,lc}^{1/3}P_0^{1/3},
\end{eqnarray}
where $R_{\rm open,lc}\equiv R_{\rm open}/R_{\rm lc}$ and $R_{\rm open}(\ge R_{\rm lc})$ is the maximum distance from the centre of the neutron star to the top of the magnetic loop on a given field line .

Using equations (\ref{sec2:number2}), (\ref{sec2:N_gamma}), (\ref{sec2:tau}), (\ref{sec2:N_p}), (\ref{sec2:gamma_s}) and (\ref{sec2:gamma_s,lt}), we obtain the upper limit on the emission location, 
\begin{eqnarray}\label{sec4:r_nB}
r_{\rm LBsyn,6}&\sim&2.5\times10^{-3}\alpha^{4/5}\xi_{-1}^{-6/5}\nu_{\rm obs,keV}^{3/5} \nonumber \\
& &\times R_{\rm open,lc}^{-1/5}E_{\rm cut,GeV}^{6/5}P_0^{-4/5}B_{\rm s,12}^{3/5}.
\end{eqnarray}

\section{Results and Discussion}

\begin{figure}
\includegraphics[width=110mm, angle=270]{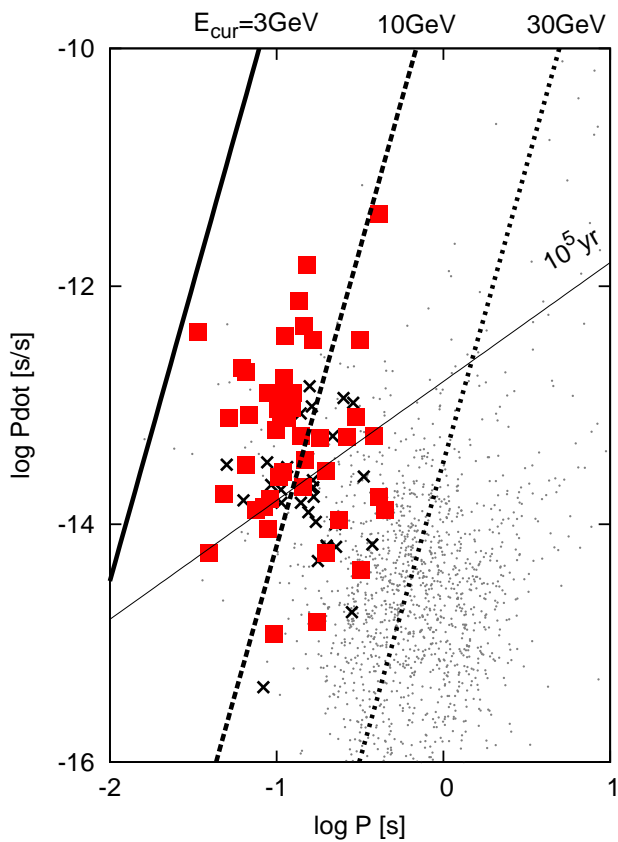}
\caption{Synchrotron radiation death lines on the $P-\dot{P}$ diagram. Thick lines are obtained from inequality (\ref{sec4:inBgammaPdotP}) for the characteristic energy of the curvature radiation $E_{\rm cur}=3$ GeV (solid line), 10 GeV (dashed line) and 30 GeV (dotted line). For other parameters, we set $\alpha=1$, $\xi=0.1$, $R_{\rm open}=R_{\rm lc}$ and $h\nu_{\rm obs}=1$ keV. Thin solid line denote the characteristic age $10^5$ yr. Large points denote the gamma-ray pulsars with non-thermal X-ray detected (red squares) and non-detected (black crosses) from 2nd Fermi Pulsar Catalog \cite{catalog}. Small dots denote other pulsars taken from ATNF Pulsar Catalog \cite{Ma05}. The characteristic energies of the curvature radiation are $\lesssim5$ GeV for middle-aged gamma-ray pulsars, so that the model does not explain their non-thermal X-ray emissions.}
\label{fig1}
\end{figure}

To explain the observed non-thermal X-ray emission, the emission location $r_{\rm sec}$ has to satisfy the condition, 
\begin{eqnarray}
r_{\rm ct}<r_{\rm sec}<r_{\rm LBsyn}.
\end{eqnarray}
Using the condition $r_{\rm ct}<r_{\rm LBsyn}$, we obtain the death lines for the synchrotron radiation on the $P$--$\dot{P}$ plane as
\begin{eqnarray}\label{sec4:inBgammaPdotP}
\dot{P}&>&0.66\alpha^{-16}\xi_{-1}^9\nu_{\rm obs,keV}^{-7} \nonumber \\
& &\times R_{\rm open,lc}^{3/2}E_{\rm cur, GeV}^{-9}P_0^5~{\rm s~s}^{-1}, 
\end{eqnarray}
where we use $B_{\rm s,12}\sim6.4P_0^{1/2}\dot{P}_{-14}^{1/2}$.

We show the results (equation \ref{sec4:inBgammaPdotP}) in figure \ref{fig1}.
Large symbols (squares and crosses) show the $\gamma$-ray pulsars from Fermi 2nd pulsar catalog \cite{catalog}. 
Red squares denote pulsars whose non-thermal X-ray emission is detected. 
This figure show that even if the luminosity ratio $\xi=0.1$ and pitch angle $\alpha=1$, the characteristic energy of the curvature radiation $E_{\rm cur}\gtrsim 5$ GeV is required to explain the observed non-thermal X-ray emission in our model. 
However, observed characteristic energy for most $\gamma$-ray pulsars typically $E_{\rm cur}\lesssim5$ GeV \cite{catalog}. 
Therefore, the proposed model of the synchrotron emission \cite[e.g., ][]{ZC97, CGZ98, CZ99} is difficult to explain the observed non-thermal component.

We briefly discuss other models.
In the synchrotron models with the photon-photon pair-production, the number density of seed photons is too small to produce the number of pairs \cite{KT14}. 
The model of the synchrotron radiation with outgoing primary particles and magnetic pair-production may be possible to explain the observed non-thermal X-ray emission (Figure 1 in \cite{KT14}). 
In this case, $\gamma$-ray photons have to be produced near the surface as the same as polar cap model \cite{DH96}.
However, the observed $\gamma$-ray pulse profile and spectral cutoff shape favor that $\gamma$-ray emission region is far from NS \cite[e.g., ][]{catalog, Pir+15}.
A possible idea to resolve this inconsistency is that more than two particle acceleration regions exist in the magnetosphere \cite[e.g., ][]{YS12, P13}.
This model should be constrained by geometrical analysis using observed pulse profiles at both $\gamma$-ray and X-ray \cite[e.g., ][]{KK11}.
Due to poor photon statistics, pulse profiles of non-thermal component have been detected for only small number of middle-aged pulsars. 
If pulse profiles will be detected for large samples in future observations such as NuSTAR and ASTRO-H, we can impose more significant constraint on the particle production in the magnetosphere.

\bigskip 
\begin{acknowledgments}
This work is supported by KAKENHI 24103006 (S.K.) and 2510447 (S.J.T.).
\end{acknowledgments}

\bigskip 

\begin{thebibliography}{99}   


\bibitem{catalog}
Abdo, A. A., et al. 2013, ApJS, 208, 17 

\bibitem{CZ99}
Cheng,~K.~S., \& Zhang,~L. 1999, ApJ, 515, 337

\bibitem{CHR86}
Cheng,~K.~S., Ho,~C., \& Ruderman,~M. 1986, ApJ, 300, 522

\bibitem{CGZ98}
Cheng,~K.~S., Gil,~J., \& Zhang,~L. 1998, ApJ, 493, L35

\bibitem{DH96}
Daugherty,~J.~K., \& Harding,~A.~K. 1996, ApJ, 458, 278

\bibitem{HR93}
Halpern,~J.~P., \& Ruderman,~M. 1993, ApJ, 157, 869

\bibitem{H13}
Hirotani,~K. 2013, ApJ, 766, 98

\bibitem{Kar+12}
Kargaltsev,~O., Durant,~M., Pavlov,~G.~G., \& Garmire,~G.~P. 2012, ApJS, 201, 37

\bibitem{KK11}
Kisaka,~S., \& Kojima,~Y. 2011, ApJ, 739, 14

\bibitem{KT14}
Kisaka,~S., \& Tanaka,~S.~J. 2014, MNRAS, 443, 2063

\bibitem{Ma05}
Manchester, R. N., Hobbs, G. B., Teoh, A., \& Hobbs, M. 2005, AJ, 129, 1993

\bibitem{MDC11}
Marelli,~M., De Luca,~A., \& Caraveo,~P.~A. 2011, ApJ, 733, 82

\bibitem{OS70}
O'Dell,~S.~L., \& Sartori,~L. 1970, ApJ, 161, L63

\bibitem{P13}
Petrova,~S.~A. 2013, ApJ, 764, 129
 
\bibitem{Pir+15}
Pierbattista,~M., Harding,~A.~K., Grenier,~I.~A., Johnson,~T.~J., Caraveo,~P.~A., Kerr,~M., \& Gonthier,~P.~L. 2015, A\&A, 575, A3

\bibitem{RD99}
Rudak,~B., \& Dyks,~J. 1999, MNRAS, 303, 477

\bibitem{RS75}
Ruderman,~M.~A., \& Sutherland,~P.~G. 1975, ApJ, 196, 51 

\bibitem{S71}
Sturrock,~P.~A. 1971, ApJ, 164, 529 

\bibitem{TC09}
Takata, J., \& Chang,~H.-K. 2009, MNRAS, 392, 400 

\bibitem{TCS08}
Takata, J., Chang,~H.-K., \& Shibata,~S. 2008, MNRAS, 386, 748 

\bibitem{TSHC06}
Takata, J., Shibata,~S., Hirotani,~K., \& Chang,~H.-K. 2006, MNRAS, 366, 1310 

\bibitem{T10}
Timokhin,~A.~N. 2010, MNRAS, 408, 2092 

\bibitem{YP04}
Yakovlev,~D.~G., \& Pethick,~C.~J. 2004, ARA\&A, 42, 169 

\bibitem{YS12}
Yuki,~S., \& Shibata,~S. 2012, PASJ, 64, 43 

\bibitem{ZC97}
Zhang,~L., \& Cheng,~K.~S. 1997, ApJ, 487, 370



\end{thebibliography}

\end{document}